\documentclass[3p,times,procedia]{elsarticle}
\usepackage{nupha_ecrc}

\volume{00}

\firstpage{1}

\journalname{Nuclear Physics A}

\runauth{B. Duclou\'e, T. Lappi, H. M\"antysaari}

\jid{nupha}

\jnltitlelogo{Nuclear Physics A}

\usepackage{amssymb}

\usepackage[figuresright]{rotating}

\usepackage{amsmath}
\usepackage{hyperref}

\def\P{{\boldsymbol P}}

\newcommand{\der}{\mathrm{d}}
\newcommand{\xt}{{{\boldsymbol x}_\perp}}
\newcommand{\yt}{{{\boldsymbol y}_\perp}}
\newcommand{\bt}{{{\boldsymbol b}_\perp}}
\newcommand{\rt}{{{\boldsymbol r}_\perp}}

\newcommand{\Pt}{{\P_\perp}}

\newcommand{\ud}{\, \mathrm{d}}
\newcommand{\tr}{\, \mathrm{Tr} \, }
\newcommand{\nc}{{N_\mathrm{c}}}

\newcommand{\Jpsi}{{J/\psi}}

\begin{document}

\begin{frontmatter}



\dochead{}

\title{Centrality dependence of forward $J/\psi$ suppression in high energy proton-nucleus collisions}


\author[jyu,hip]{B. Duclou\'e}
\author[jyu,hip]{T. Lappi}
\author[bnl]{H. M\"antysaari}
\address[jyu]{Department of Physics, University of Jyv\"askyl\"a, P.O. Box 35, 40014 University of Jyv\"askyl\"a, Finland}
\address[hip]{Helsinki Institute of Physics, P.O. Box 64, 00014 University of Helsinki, Finland}
\address[bnl]{Physics Department, Brookhaven National Laboratory, Upton, NY 11973, USA}

\begin{abstract}
The production of forward $J/\psi$ mesons in proton-nucleus 
collisions can provide important information on gluon saturation. In a previous work we studied this process in the Color Glass Condensate framework, describing the target using a dipole cross section fitted to HERA inclusive data and extrapolated to the case of a nuclear target using the optical Glauber model.
In this work we study the centrality dependence of the nuclear suppression in this model and compare our results with recent LHC data for this observable.
\end{abstract}


\end{frontmatter}


\section{Introduction}

The production of particles at forward rapidity is an important tool to study gluon saturation since it probes the target at very small $x$. In particular, the charm quark mass being of the same order of magnitude as the saturation scale, forward $J/\psi$ production should be sensitive to these dynamics. In addition, the charm quark mass is also large enough to provide a hard scale, making a perturbative treatment possible. On the experimental side, $J/\psi$ production has been the subject of many studies, providing a lot of data to confront theory with experiment.

In a recent work~\cite{Ducloue:2015gfa} we studied, in the Color Glass Condensate (CGC) framework, the production of forward $J/\psi$ mesons in proton-proton and proton-nucleus collisions at the LHC. We showed that, when using the Glauber approach to generalize the dipole cross section of a proton to the one of a nucleus, the nuclear suppression for minimum bias events is smaller than in previous CGC calculations such as~\cite{Fujii:2013gxa} and much closer to experimental data\footnote{The authors of~\cite{Fujii:2013gxa} have recently presented updated results~\cite{Fujii:2015lld} which are similar to those obtained in~\cite{Ducloue:2015gfa}.}. Here we will study the centrality dependence of the nuclear suppression in this model. This observable was recently studied at the LHC by the ALICE Collaboration~\cite{Adam:2015jsa}.

\section{Formalism}

The formalism for gluon and quark pair production in the dilute-dense limit of the CGC has been studied in detail in Refs.~\cite{Blaizot:2004wu,Blaizot:2004wv} (see also Ref.~\cite{Kharzeev:2012py}) and used in several works, such as~\cite{Fujii:2005rm,Fujii:2006ab,Fujii:2013gxa,Fujii:2013yja,Ma:2015sia}. This allows to compute the cross section for $c\bar{c}$ pair production, which can be found in Ref.~\cite{Fujii:2013gxa} and is the key ingredient to study $J/\psi$ production. To describe the hadronization of the $c\bar{c}$ pairs into $J/\psi$ mesons, we will use the simple color evaporation model in which a fixed fraction of the $c\bar{c}$ pairs produced below the $D$-meson threshold in either the color singlet or octet state is assumed to hadronize into $\Jpsi$ mesons. The differential cross section with respect to the transverse momentum $\P_{\perp}$ and the rapidity $Y$ of the produced $\Jpsi$ then reads
\begin{align} 
	\frac{\ud\sigma_{\Jpsi}}{\ud^2\P_{\perp}\ud Y}
	=
	F_{\Jpsi} \; \int_{4m_c^2}^{4M_D^2} \ud M^2
	\frac{\ud\sigma_{c\bar c}}
	{\ud^2\P_{\perp} \ud Y \ud M^2}
	\, ,
	\label{eq:dsigmajpsi}
\end{align}
where $m_c$ is the charm quark mass, $m_D=1.864$ GeV is the $D$ meson mass and $\frac{\ud\sigma_{c\bar c}}{\ud^2\P_{\perp} \ud Y \ud M^2}$ is the cross section for $c\bar{c}$ pair production with transverse momentum $\P_{\perp}$, rapidity $Y$ and invariant mass $M$. The nonperturbative constant $F_{\Jpsi}$ is related to the probability for a $c\bar{c}$ pair with an invariant mass between $2m_c$ and $2M_D$ to transition to a $\Jpsi$. In this work we will focus on the nuclear modification factor for which the value of $F_{\Jpsi}$ plays no role since it is assumed to be the same in proton-proton and proton-nucleus collisions.

Since we work at forward rapidity, the projectile proton is probed at large $x$ and therefore the gluon density inside it can be described using collinear factorization. For this we use the MSTW 2008~\cite{Martin:2009iq} parametrization at leading order since the remaining of our calculation is done at this order.
The target, on the other hand, is probed at very small $x$ and the information about its gluon density is encoded in the function 
\begin{equation}
	S_{_Y}(\xt-\yt) = \frac{1}{\nc }\left< \tr U^\dag(\xt)U(\yt)\right>,
\end{equation}
where $U(\xt)$ is a fundamental representation Wilson line in the color field of the target. $S_{_Y}(\rt)$ is obtained by solving numerically the running coupling Balitsky-Kovchegov equation~\cite{Balitsky:1995ub,Kovchegov:1999ua,Balitsky:2006wa}. In the case of a proton target we use the MV$^e$ parametrization introduced in Ref.~\cite{Lappi:2013zma} as the initial condition. It involves, similarly to the AAMQS~\cite{Albacete:2010sy} one, several parameters which are fit to HERA DIS data~\cite{Aaron:2009aa}. In this case there is no explicit dependence on the impact parameter.

In the case of a nuclear target we use, again following~\cite{Lappi:2013zma}, the optical Glauber model to relate the dipole amplitude of a nucleus to the one of a proton. The only additional input needed compared to the proton case is the standard Woods-Saxon distribution $T_A(\bt)$, where $\bt$ is the impact parameter:
\begin{equation}
T_A(\bt)= \int \der z \frac{n}{1+\exp \left[ \frac{\sqrt{\bt^2 + z^2}-R_A}{d} \right]} \; ,
\end{equation}
with $d=0.54\,\mathrm{fm}$, $R_A=(1.12A^{1/3}-0.86A^{-1/3})\,\mathrm{fm}$ and $n$ is fixed such that $\int \der^2\bt T_A(\bt)=1$.
This model has an explicit impact parameter dependence, which we will relate to the centrality classes used by the experiments in the following way. The experimental results are reported in centrality classes together with an estimate of the mean number of collisions in each class, $\langle N_\text{coll} \rangle_\text{exp.}$. We take this estimate and translate it into a value of the impact parameter in the optical Glauber model using the relation $N_\text{coll opt.}(\bt)=\langle N_\text{coll} \rangle_\text{exp.}$, where $N_\text{coll opt.}$ is given by
\begin{equation}
N_\text{coll opt.}(\bt) = A T_A(\bt) \sigma_\text{inel} \; ,
\end{equation}
with $\sigma_\text{inel}$ being the total inelastic nucleon-nucleon cross section. The centrality dependent results presented in the following section are obtained with this fixed impact parameter in each centrality class. Using instead distributions in the impact parameter space would be more consistent and could affect our results significantly depending on the exact shape of these distributions but this would require experimental access not only to $\langle N_\text{coll} \rangle$ but to the full $N_\text{coll}$ distribution in each centrality class.

\section{Results}

\begin{figure}[h]
	\def\picheight{4cm}
	\centering\includegraphics[height=\picheight]{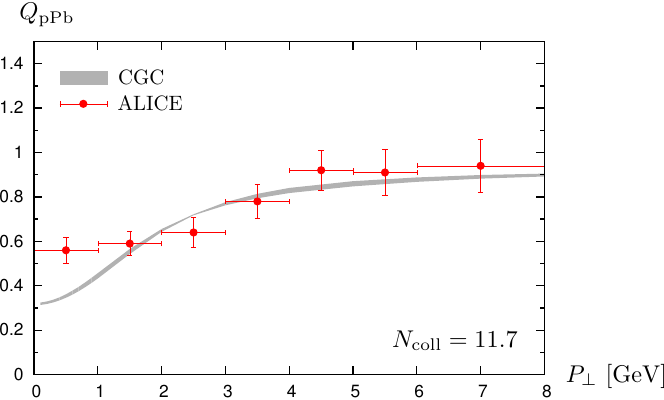}
	\hspace{0.45cm}
	\includegraphics[height=\picheight]{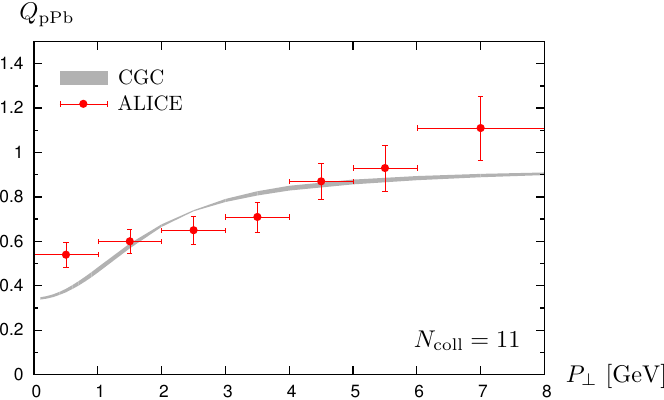}
	
	\vspace{0.2cm}
	\includegraphics[height=\picheight]{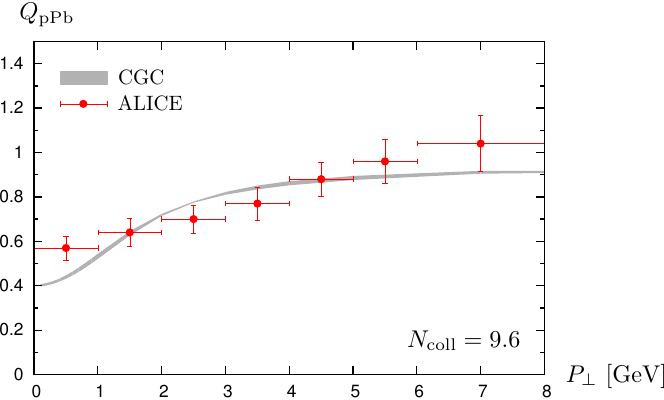}
	\hspace{0.4cm}
	\includegraphics[height=\picheight]{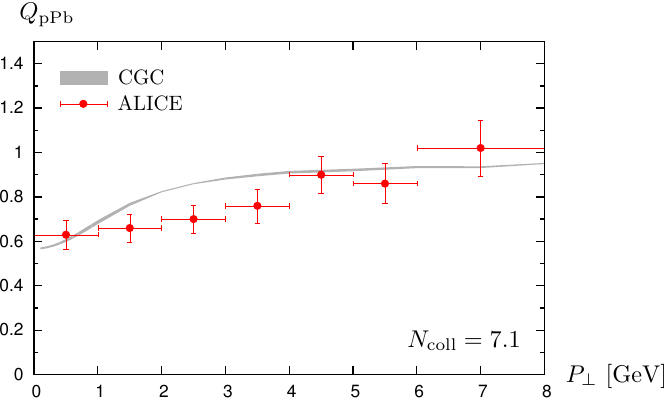}
	
	\vspace{0.2cm}
	\includegraphics[height=\picheight]{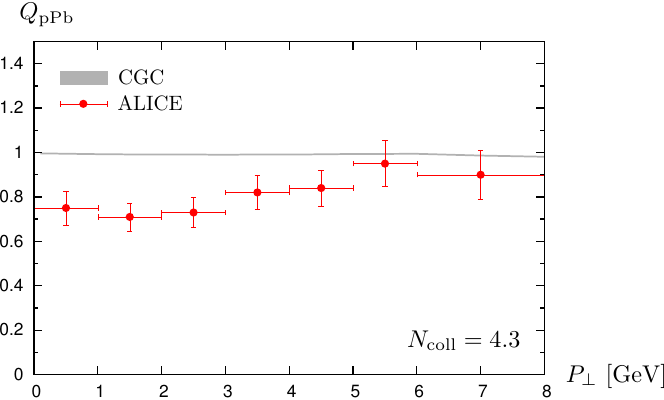}
	\caption{Nuclear modification factor $Q_\text{pPb}$ as a function of $P_\perp$ in different centrality bins compared with ALICE data~\cite{Adam:2015jsa}.}
	\label{fig:QpA_bins}
\end{figure}

The nuclear modification factor of forward $\Jpsi$ production in proton-lead collisions at the LHC was recently measured by the ALICE Collaboration in several centrality classes at $\sqrt{s_{NN}}=5$ TeV~\cite{Adam:2015jsa}. As explained previously, to compare our results with these data we fix the impact parameter $\bt$ such that $N_\text{coll. opt}(\bt)=\langle N_\text{coll} \rangle_\text{ALICE}$ in each centrality class considered by ALICE. Note that we do not consider the most peripheral class because this procedure would lead to an impact parameter value for which the saturation scale of the nucleus falls below the one of the proton. In Fig.~\ref{fig:QpA_bins} we show the results of our calculation compared with ALICE data for the nuclear modification factor $Q_\text{pPb}$, defined as
\begin{equation}
Q_{\rm pPb}= \frac{\frac{\ud N^\text{pPb}}{\ud^2 \Pt \ud Y}}
{ A \langle T_A \rangle \frac{\ud\sigma^\text{pp}}{\ud^2 \Pt \ud Y}} \; ,
\end{equation}
as a function of $P_\perp$ for the five most central classes. The uncertainty band, which is quite narrow on this plot, contains the variation of the charm quark mass between 1.2 and 1.5 GeV and of the factorization scale between $M_\perp/2$ and $2M_\perp$ with $M_\perp=\sqrt{M^2+P_{\perp}^2}$ where $M$ is the $c\bar{c}$ pair's invariant mass. We observe that the description of the data is quite good for the three most central bins, but in the optical Glauber model $Q_\text{pPb}$ approaches unity too quickly with deacreasing $N_\text{coll}$. This is clearly visible for the fifth bin where $Q_\text{pPb}$ is almost constant and very close to unity in our calculation while the data still shows a significant variation with $P_\perp$.

\section{Conclusions}

In this work we have investigated the centrality dependence of $J/\psi$ suppression in high energy proton-nucleus collisions. We have used the same framework as in Ref.~\cite{Ducloue:2015gfa}, where it was shown to lead to a nuclear modification factor in minimum bias collisions much closer to experimental data than previous calculations in the Color Glass Condensate. However this model seems to predict a too strong centrality dependence compared with recent ALICE data~\cite{Adam:2015jsa}. Nevertheless, one should keep in mind that here we have estimated the suppression in each centrality class considered in Ref.~\cite{Adam:2015jsa} by using a fixed impact parameter corresponding to the average number of binary collisions estimated by ALICE. For a more consistent comparison, one should instead consider distributions in the impact parameter space. We leave this issue for future work.

\section*{Acknowledgments}
T.~L. and B.~D. are supported by the Academy of Finland, projects
267321 and 273464. 
H.~M. is supported under DOE Contract No. DE-SC0012704.
This research used computing resources of 
CSC -- IT Center for Science in Espoo, Finland.
We would like to thank C. Hadjidakis and I. Lakomov for 
discussions on the ALICE data.

\end{document}